\PassOptionsToPackage{table}{xcolor} 
\documentclass{article} 
\usepackage[submission]{template/colm2025_conference} 

\usepackage{graphicx}       
\usepackage{microtype}      
\usepackage{hyperref}       
\usepackage{url}            
\usepackage{amsmath}        
\usepackage{amssymb}        
\usepackage{enumitem}       
\usepackage{xspace}         

\usepackage{booktabs}       
\usepackage{array}          
\usepackage{tabularx}       
\usepackage{longtable}      
\usepackage{siunitx}        
\usepackage{subfig}         

\usepackage{tcolorbox}      
\usepackage{xcolor}  
\usepackage{multirow,makecell}       
\usepackage{wrapfig}        
\usepackage{lineno}         
\usepackage{lipsum}         
\usepackage{nicematrix}     

\hypersetup{colorlinks,citecolor={blue}}

\usepackage{listings}
\usepackage{listings-ext}
\definecolor{delim}{RGB}{20,105,176}
\definecolor{numb}{RGB}{106, 109, 32}
\definecolor{string}{rgb}{0.64,0.08,0.08}

\lstdefinelanguage{json}{
    numbers=left,
    numberstyle=\small,
    frame=single,
    rulecolor=\color{black},
    showspaces=false,
    showtabs=false,
    breaklines=true,
    postbreak=\raisebox{0ex}[0ex][0ex]{\ensuremath{\color{gray}\hookrightarrow\space}},
    breakatwhitespace=true,
    basicstyle=\ttfamily\small,
    upquote=true,
    morestring=[b]",
    stringstyle=\color{string},
    literate=
     *{0}{{{\color{numb}0}}}{1}
      {1}{{{\color{numb}1}}}{1}
      {2}{{{\color{numb}2}}}{1}
      {3}{{{\color{numb}3}}}{1}
      {4}{{{\color{numb}4}}}{1}
      {5}{{{\color{numb}5}}}{1}
      {6}{{{\color{numb}6}}}{1}
      {7}{{{\color{numb}7}}}{1}
      {8}{{{\color{numb}8}}}{1}
      {9}{{{\color{numb}9}}}{1}
      {\{}{{{\color{delim}{\{}}}}{1}
      {\}}{{{\color{delim}{\}}}}}{1}
      {[}{{{\color{delim}{[}}}}{1}
      {]}{{{\color{delim}{]}}}}{1},
}

\newcommand{\ourdataset}{\textsc{OpenCodeInstruct}\xspace}
\newcommand{\opentag}[1]{\textless#1\textgreater}
\newcommand{\closetag}[1]{\textless/#1\textgreater}

\title{OpenCodeInstruct: A Large-scale Instruction Tuning Dataset for Code LLMs}

\author{Wasi Uddin Ahmad, Aleksander Ficek, Mehrzad Samadi,  \\
{\bf Jocelyn Huang, Vahid Noroozi, Somshubra Majumdar, Boris Ginsburg} \\ [1pt]
NVIDIA \\ [1pt]
Santa Clara, CA 95051, USA \\ [1pt]
\texttt{\{wasiuddina, smajumdar, vnoroozi, aficek\}@nvidia.com} \\ [1pt]
\texttt{\url{https://huggingface.co/datasets/nvidia/OpenCodeInstruct}} 
}

\colmsubmissionfalse\colmpreprinttrue\colmfinalfalse

\begin{document}

\ifcolmsubmission
\linenumbers
\fi

\maketitle

\begin{abstract}

Large Language Models (LLMs) have transformed software development by enabling code generation, automated debugging, and complex reasoning. However, their continued advancement is constrained by the scarcity of high-quality, publicly available supervised fine-tuning (SFT) datasets tailored for coding tasks. To bridge this gap, we introduce \ourdataset, the largest open-access instruction tuning dataset, comprising 5 million diverse samples. Each sample includes a programming question, solution, test cases, execution feedback, and LLM-generated quality assessments. We fine-tune various base models, including LLaMA and Qwen, across multiple scales (1B+, 3B+, and 7B+) using our dataset. Comprehensive evaluations on popular benchmarks (HumanEval, MBPP, LiveCodeBench, and BigCodeBench) demonstrate substantial performance improvements achieved by SFT with \ourdataset. We also present a detailed methodology encompassing seed data curation, synthetic instruction and solution generation, and filtering. 

\end{abstract}

\section{Introduction}

Large language models (LLMs), pre-trained on trillions of code tokens, have achieved remarkable success across a broad spectrum of software engineering tasks \citep{hui2024qwen2, guo2024deepseek, wu2024repoformer,
xia2023automated, shypula2023learning, athiwaratkun2023multilingual, chen2021evaluating, austin2021program, chen2021evaluating, lachaux2020unsupervised}.
To enhance their ability to follow natural language instructions and tackle more complex development scenarios, these models are often further refined through instruction tuning, a process that aligns model outputs with user intent using curated instruction-response pairs \citep{jimenez2023swe, mundler2024swtbench, miserendino2025swe}. High-quality instruction-following datasets play a critical role in this stage, enabling LLMs to better bridge the gap between natural language and executable code.

Generating high-quality instruction data for fine-tuning large language models (LLMs) is a challenging and resource intensive task. Human annotation, as exemplified by the large-scale dataset used to train Llama-3 \citep{ouyang2022training, grattafiori2024llama}, can yield high-quality results but is often prohibitively expensive. It has led to widespread adoption of knowledge distillation techniques using synthetic data generation \citep{gunasekar2023textbooks, wei2023magicoder, yu-etal-2024-wavecoder, zheng-etal-2024-opencodeinterpreter, majumdar2024genetic}. One influential line of work includes \textsc{Self-Instruct} \citep{wang-etal-2023-self-instruct} and \textsc{Evol-Instruct} \citep{xu2024wizardlm}, which generate instruction data via in-context learning entirely from limited access to external data. Another emerging approach, \textsc{OSS-Instruct} \citep{wei2023magicoder}, constructs instruction data by leveraging real-world code snippets and generating corresponding prompts \citep{wei2024selfcodealign}. While more cost effective, these approaches often require access to proprietary models and data. Unlike many high-performing LLMs for code that do not disclose their instruction tuning methodologies or datasets, \citep{guo2024deepseek, grattafiori2024llama, hui2024qwen2}, \citet{huang2024opencoder} released a fully open-source coding LLM, including its pretraining and supervised fine-tuning datasets. Their SFT dataset, comprising 435k examples, represents a significant increase over the previously largest publicly available code instruction corpus.



\begin{table}[t]
\centering
\def\arraystretch{1.1}%
\begin{tabular}{l r}
\toprule
Datasets & \# Sample \\
\midrule
CodeAlpaca \citep{codealpaca}                                           & 20,000 \\
CodeSeaXDataset \citep{yu-etal-2024-wavecoder}                          & 20,000 \\
SelfCodeAlign \citep{wei2024selfcodealign}      & 50,000 \\
Evol-Instruct-Code-80k-v1 \citep{wizardcoder_data}                      & 80,000 \\
Magicoder-OSS-Instruct \citep{wei2023magicoder}                         & 75,000 \\
Magicoder-Evol-Instruct \citep{wei2023magicoder}                        & 110,000 \\
OpenCoder-LLM-sft-stage2 \citep{huang2024opencoder}                     & 435,000 \\
{\bf \ourdataset}                                              & {\bf 5,000,000} \\
\bottomrule
\end{tabular}
\caption{\ourdataset vs. other publicly available code-instruction tuning datasets.}
\label{tab:data_stat}
\end{table}

We present \ourdataset, the most extensive code instruction dataset (in Python) created to date (see comparison in \autoref{tab:data_stat}), designed to facilitate instruction tuning of large language models and accelerate advancements in code LLM research. Unlike previous approaches that relied on limited seed instructions or code snippets, \ourdataset leverages a significantly larger and more diverse seed set. It leverages 1.43 million general coding instructions (derived from Python functions extracted from 
GitHub using \textsc{OSS-Instruct}) and 25,443 algorithmic questions from TACO \citep{li2023taco} as seeds, resulting in a comprehensive synthetic dataset of 5 million samples for instruction tuning. \ourdataset employs a scalable synthetic data generation framework \citep{majumdar2024genetic}, integrating the strengths of \textsc{Self-Instruct} and \textsc{Evol-Instruct} to further enhance data quality. Additionally, it incorporates LLM-generated unit tests for feedback aggregation and LLM judgment for sample quality assessment.

Using \ourdataset, we fine-tuned \emph{base} LLMs -- Llama3 \citep{grattafiori2024llama} and Qwen2.5-Coder \citep{hui2024qwen2} across different parameter scales: 1B+, 3B+, and 7B+. Our fine-tuned models, OCI-Llama3 and OCI-Qwen2.5-Coder, demonstrated a substantial performance gain over their instruction-tuned counterparts, Llama3-Instruct and Qwen2.5-Coder-Instruct. Moreover, we conducted comprehensive ablation and analysis with several key findings: (1) Fine-tuning with just 500k samples from \ourdataset surpassed the original Llama-3 and Qwen2.5-Coder instruct models, with further fine-tuning yielding further gains; (2) LLM judgment proved to be a more effective indicator of instruction quality than execution-based feedback; (3) Genetic-Instruct, which integrates both Evol-Instruct and Self-Instruct, yielded higher performance compared to using instructions generated by either approach alone; (4) Larger seed sets for synthetic data generation improved downstream code generation; (5) Both generic and algorithmic coding instructions contributed positively as seeds; and (6) Natural language to code (NL-to-Code) instruction formatting significantly outperformed code-to-code style prompting (as used in HumanEval).

The contributions of this work can be summarized as follows:
\begin{enumerate}[leftmargin=*]
    \item {\bf Advancement of Code Instruction Tuning}: We present \ourdataset
    , the largest publicly available code instruction tuning dataset to date, comprising 5 million samples with rich metadata (unit tests, execution feedback, LLM judgments), significantly expanding the resources available for code instruction tuning. 
    
    \item {\bf Demonstrated Performance Gains}: Fine-tuning Llama3 and Qwen2.5-Coder with \ourdataset yields substantial performance improvements over their instruction-tuned counterparts on key code generation benchmarks, including HumanEval, MBPP, LiveCodeBench, and BigCodeBench.

    \item {\bf In-depth Analysis and Valuable Research Insights}: Extensive ablation and analyses reveal key findings on data scaling, generation techniques, seed sets, and instruction formatting, guiding future research in the field.
\end{enumerate}

\begin{figure*}[ht]
\centering
\includegraphics[width=\textwidth]{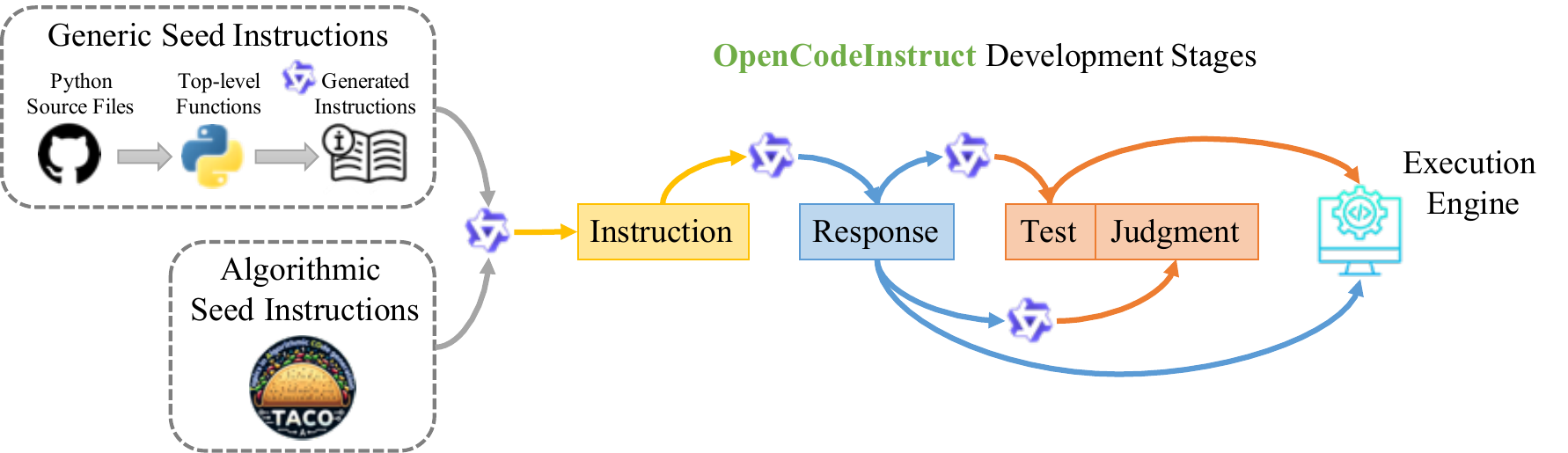}
\caption{Overview of the \ourdataset development stages.}
\label{fig:oci_overview}
\end{figure*}

\section{\ourdataset: Large-scale Coding Instruction Tuning Dataset}
The \ourdataset development stages are illustrated in \autoref{fig:oci_overview}. \ourdataset uses two main sets of coding instruction collections as the initial seeds: a large-scale generic one generated synthetically, and a small-scale algorithmic set of non-synthetic coding problems. The large-scale seed instructions are generated by using \textsc{OSS-Instruct} algorithm \citep{wei2023magicoder} based on a set of Python functions extracted from Github. This collection covers a wide range of coding problems, while the smaller scale collection is a high-quality set of questions focused on algorithmic coding problems. Then, \ourdataset uses a scalable synthetic data generation framework, \textsc{Genetic-Instruct} \citep{majumdar2024genetic} to generate synthetic coding instructions, and their corresponding responses. We further augments synthetic data samples with unit tests, execution feedback, and LLM judgment on quality and correctness. In the following sections, we provide detailed explanations of these steps.

\subsection{Creation of the Initial Seed Collection}
Previous research has shown that the quality of synthetic data is highly dependent on both the generator LLM's performance and the initial seed set. Small seed sets and weaker generator LLMs often lead to duplicate instruction instances, reducing instruction tuning effectiveness  \citep{yan2024understanding, lee-etal-2022-deduplicating, xu2022learning}. To address this, we employed the following two main set of initial seeds in the \ourdataset pipeline in parallel to enhance the diversity and widen the range of the domains covered by the generated instructions. The \textsc{Genetic-Instruct} framework has a deduplication process based on n-grams which prevents instruction duplication.

\begin{wraptable}{r}{55mm}
\centering
\vspace{-3mm}
\def\arraystretch{1.0}%
\resizebox{55mm}{!}{
\begin{tabular}{l | r}
\toprule
Source & \# Questions \\
\midrule
AIZU & 2151 \\
AtCoder & 1440 \\
CodeChef & 3352 \\
CodeForces & 8193 \\
Codewars & 2460 \\
GeeksForGeeks & 2680 \\
HackerEarth & 2390 \\
HackerRank & 764 \\
Kattis & 1236 \\
LeetCode & 777 \\
\midrule
Total & 25,443 \\
\bottomrule
\end{tabular}
}
\vspace{-2mm}
\caption{Question distribution in TACO \citep{li2023taco} across various competitive coding platforms.}
\label{tab:taco_stat}
\vspace{-10mm}
\end{wraptable}



\paragraph{Small-scale algorithmic coding questions} We leverage 25,443 algorithmic questions from TACO \citep{li2023taco} as seed instructions. \autoref{tab:taco_stat} shows the question distribution collected from various competitive coding platforms. These questions, covering diverse data structures and algorithms, enrich the diversity of the synthetic generated instructions.

\paragraph{Large-scale generic coding instructions} 
To build this set, we collected a set of Python functions from GitHub, following the data collection pipeline outlined in \citet{wei2024selfcodealign}. It involved extracting Python functions with docstrings, followed by a rigorous filtering process: type checking with Pyright, removal of benchmark items, elimination of poorly documented functions, and deduplication. Using the collected seed functions, we employed the \textsc{OSS-Instruct} framework \citet{wei2023magicoder} to generate diverse instructions. Specifically, we prompted the Qwen2.5-32B-Instruct model to create a coding task inspired by each one of the Python functions. This process resulted in 1.43 million coding instructions, which were subsequently used as seed questions for the \ourdataset pipeline. It is important to note that while OSS-Instruct generates both coding instructions and solution code, we only utilized the generated instructions as seeds, discarding the solution code. 

\subsection{Instruction Generation}
\ourdataset adopts the \textsc{Genetic Instruct} framework \citep{majumdar2024genetic} that begins with a set of initial instructions and employs LLMs to generate instructions and their corresponding code solutions through two evolutionary operations: mutation and crossover that mimics \textsc{Evol-Instruct} \citep{luo2024wizardcoder} and \textsc{Self-Instruct} \citep{wang-etal-2023-self-instruct}, respectively. In the mutation operation, LLM generates a new instruction given an input instruction and a specific task. The task is chosen randomly from a set of five tasks introduced in \citet{luo2024wizardcoder}. In the crossover operation, an Instruct-LLM is prompted to generate multiple diverse set of new instructions based on a given set of instructions from the seeds. Despite \textsc{Genetic-Instruct}'s iterative nature, we ran it for a single generation, generating nearly 9 million synthetic instructions. We refer the readers to \citet{majumdar2024genetic} for further details about \textsc{Genetic-Instruct}.

\subsubsection{Data Cleaning and Decontamination}
While the \textsc{Genetic-Instruct} framework inherently deduplicates the generated instructions, we further refined the dataset with the following two steps:

\begin{itemize}[leftmargin=*]
    \item {\bf Filtering}: We filter out instructions that include Python code snippets because we observed that they are significantly noisy and primarily created due to one of the tasks in \textsc{Evol-Instruct} pertaining to code repair/refactoring. Moreover, those instructions were deemed unhelpful for our target code generation tasks.
    \item {\bf Decontamination}: We used an n-gram-based decontamination method to remove any overlap between our instructions and the evaluation benchmarks.\footnote{\url{https://github.com/huggingface/open-r1/blob/main/scripts/decontaminate.py}} 
\end{itemize}

Following data cleaning and decontamination, we retained approximately 5 million synthetic coding questions that we use for response generation in the next step.

\subsection{Response Generation}
Subsequently, we generated the answers for the generated instructions which are supposed to include the coding solution to the problems. To generate high-quality code solutions, we prompted the Qwen2.5-Coder-32B-Instruct model with the instructions and asked it to provide the solution. Additionally, to analyze the impact of the coder LLM, we generated code solutions using Qwen2.5-32B-Instruct and QwQ-32B-Preview as well.

\paragraph{What skills are used or demonstrated in responses?}
To analyze the coding skills relevant to the instructions and responses, \ourdataset includes a list of coding skills generated automatically by LLMs as metadata. We prompted the Qwen2.5-32B-Instruct model to select three skills which are covered by a code solution from a predefined list (\autoref{fig:skills_prompt}). However, the model sometimes generated skills outside this list, reflecting the broader relevance to the instruction and code. A word cloud visualization of these skills is presented in \autoref{plot:skills}, demonstrating a broad range of data structure and algorithmic concepts are covered in \ourdataset.

\subsection{Test Case Generation and Execution}
\label{sec:test_case_generation}
To broaden the applications of the our dataset, we followed the methodology of \citet{ficek2025scoringverifiersevaluatingsynthetic} and generated 10 assertion-style unit tests for each question-solution pair using Qwen2.5-Coder-32B-Instruct (prompt in \autoref{fig:utg_prompt}). One important usage of unit tests is in reinforcement learning (RL) with execution feedback which has gained popularity in enabling reasoning capability in LLMs recently \cite{guo2025deepseek}. After generating the test cases, we executed all the solutions on their corresponding generated unit tests and included the results along with the pass rate for each solution as metadata. 

\begin{wrapfigure}{r}{0.45\textwidth}
    \vspace{-4mm}
    \centering
    \includegraphics[width=0.4\textwidth]{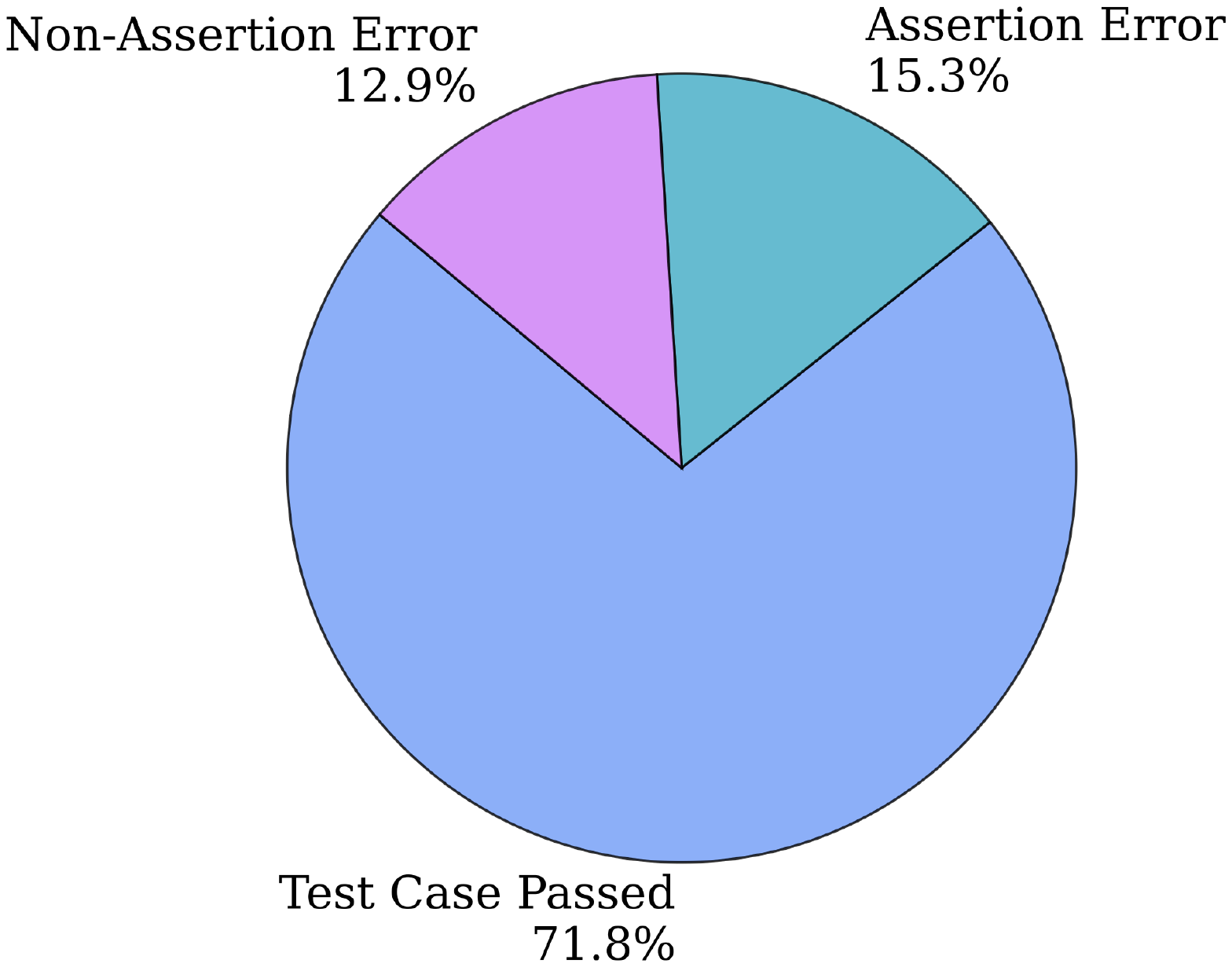}
    \vspace{-2mm}
    \caption{Unit tests pass/fail rates for \ourdataset samples.}
    \label{plot:oci_error_distribution}
    \vspace{-8mm}
\end{wrapfigure}
We showed the test case pass/fail distributions and detailed error categorizations in \autoref{plot:oci_error_distribution} and \autoref{plot:oci_error_type_distribution} ,respectively. \autoref{plot:ut_score_distribution} shows the unit test pass rate of all of the samples. Generally, solutions are skewed towards pass rates of 1.0 and 0.0, demonstrating a bimodal distribution. The high frequency of solutions that pass all tests can be explained by self-consistency bias in the models, where if they generate a solution they are also likely to believe the solution is correct \citep{huang2024largelanguagemodelsselfcorrect}. The high number of completely failing solutions can be attributed to incorrect test cases or practically un-executable solutions due to, for example, timeout errors.


\subsection{Response Quality Assessment}
\label{sec:response_quality}


To automate response quality assessment, \ourdataset utilizes the LLM-as-a-judge approach, which is based on the established competence of LLMs in matching human preferences \citep{zheng2023judging}. We prompted Qwen2.5-Coder-32B-Instruct to assess each solution's requirement conformance, logical correctness, and edge case consideration (prompt shown in \autoref{fig:llm_judge_prompt}). We included the assessment scores along with their justifications in the dataset as metadata. We averaged the three assessment scores and displayed their distribution in \autoref{plot:llm_score_distribution}. Consistent with unit test generation, the model generally rated the provided solutions highly, demonstrating self-consistency bias. The slightly increased presence of samples with an average score of 1.0 is likely attributable to a small subset of entirely incorrect or incoherent solutions. 


\section{Main Evaluation}
For our evaluation, we selected Llama3 and Qwen2.5-Coder as our \emph{base} LLMs, fine-tuning their 1B+, 3B+, and 7B+ variants using \ourdataset. We trained these models for 3 epochs on NVIDIA A100-80GB GPUs, employing an initial learning rate of $5e-6$ with 100 warmup steps and a CosineAnnealing scheduler. The AdamW optimizer \citep{kingma2014adam} was used with a batch size of 2048 and a maximum sequence length of 2048. The final models were generated by averaging checkpoints saved at the end of each epoch. We utilized tensor parallelism and BF16 precision to accelerate the training process. The main evaluation results are presented in \autoref{tab:main}. As baselines, we compared our fine-tuned models with their instruction-tuned versions and also the OpenCoder models which are trained on the largest publicly available instruction tuning datasets for coding. 

\paragraph{HumanEval and MBPP}
We reported the evaluations on HumanEval \citep{chen2021evaluating}, MBPP \citep{austin2021program}, HumaneEval+ \citep{liu2023is}, and MBPP+ \citep{liu2023is} which are the most common benchmarks for function-level code generation. The results indicate that \ourdataset substantially improves Llama3 performance and our models significantly exceed their instruction-tuned counterparts significantly, possibly due to its non-code-specific training. In contrast, for Qwen2.5-Coder, a specialized code LLM, fine-tuning with our dataset resulted in scores that were either competitive with or exceeded its instruction-tuned counterparts.

\paragraph{LiveCodeBench} LiveCodeBench \citep{jain2025livecodebench} is an extensive, contamination-free benchmark created to assess the coding capabilities of LLMs. It provides a continuously updated and diverse set of challenges by systematically collecting new problems from leading competitive programming platforms, such as LeetCode\footnote{\url{https://leetcode.com}}, AtCoder\footnote{\url{https://atcoder.jp}}, and CodeForces\footnote{\url{https://codeforces.com}}.  In this work, we use LiveCodeBench-\emph{v4}, comprising 713 coding problems. Our evaluation demonstrates that finetuning with \ourdataset significantly enhances Qwen2.5-Coder models. However, the performance improvements for smaller Llama3 models (1B+ and 3B+) are marginal, likely due to the complexity of LiveCodeBenchmark samples, which may require LLMs larger than 7B to effectively solve them.

\paragraph{BigCodeBench-Instruct}
BigCodeBench-Instruct, a natural language instruction adaptation of BigCodeBench \citep{zhuo2025bigcodebench}, challenges LLMs with complex function calling tasks. The dataset contains 1,140 tasks, each with 5.6 test cases, requiring the use of multiple function calls from 139 libraries across 7 domains. Evaluation results indicate that finetuning with \ourdataset results in better performance than their instruction-tuned counterparts for both evaluated models, particularly in the 3B+ and 7B+ size ranges.

\begin{table}[t]
\centering
\resizebox{1.0\linewidth}{!} {%
\def\arraystretch{1.2}%
{\setlength{\tabcolsep}{4pt}
\begin{NiceTabular}{l |c c|c c|c c}
\toprule
\multirow{2}{*}{\bf Model} & \multicolumn{2}{c|}{\bf HumanEval} & \multicolumn{2}{c|}{\bf MBPP} & {\bf LiveCodeBench} & {\bf BigCodeBench} \\
& HE & HE+ & MBPP & MBPP+ & Avg & Full \\ 
\midrule
\multicolumn{7}{c}{\bf 1B+ Models} \\
\midrule
Llama-3.2-1B-Instruct                       & 29.3 & 26.8 & 40.2 & 34.1 & 4.5 & 8.1 \\
Qwen2.5-Coder-1.5B-Instruct                 & 70.7 & 66.5 & 69.2 & 59.4 & 14.6 & 32.5 \\
OpenCoder-1.5B-Instruct                     & 72.5 & 67.7 & 72.7 & 61.9 & 12.8 & 33.3 \\
\rowcolor{gray!20}OCI-Llama-3.2-1B          & 51.8 & 50.0 & 53.4 & 46.6 & 4.6 & 8.5 \\
\rowcolor{gray!20}OCI-Qwen2.5-Coder-1.5B    & {\bf 78.7} & {\bf 73.8} & {\bf 80.2} & {\bf 68.3} & {\bf 25.7} & {\bf 33.8} \\
\midrule
\multicolumn{7}{c}{\bf 3B+ Models} \\
\midrule
Llama-3.2-3B-Instruct                       & 50.0 & 45.7 & 57.1 & 48.1 & 13.2 & 21.9 \\
Qwen2.5-Coder-3B-Instruct                   & 84.1 & {\bf 80.5} & 73.6 & 62.4 & 23.7 & 35.8 \\
\rowcolor{gray!20}OCI-Llama-3.2-3B          & 68.9 & 65.2 & 69.8 & 61.1 & 13.5 & 26.2 \\
\rowcolor{gray!20}OCI-Qwen2.5-Coder-3B      & {\bf 84.8} & 79.7 & {\bf 81.0} & {\bf 69.3} & {\bf 31.1} & {\bf 38.1} \\
\midrule
\multicolumn{7}{c}{\bf 7B+ Models} \\
\midrule
Llama-3.1-8B-Instruct                       & 69.5 & 62.8 & 68.3 & 60.6 & 19.2 & 33.6 \\
Qwen2.5-Coder-7B-Instruct                   & {\bf 88.4} & {\bf 84.1} & 83.5 & 71.7 & 32.3 & 41.0 \\
OpenCoder-8B-Instruct                       & 83.5 & 78.7 & 79.1 & 69.0 & 23.2 & 40.3 \\
\rowcolor{gray!20}OCI-Llama-3.1-8B          & 78.7  & 73.2  & 77.5  & 66.4  & 24.1  & 37.1 \\
\rowcolor{gray!20}OCI-Qwen2.5-Coder-7B      & 87.8  & {\bf 84.1}  & {\bf 86.8}  & {\bf 74.9}  & {\bf 39.7} & {\bf 43.6} \\
\bottomrule
\end{NiceTabular}
}}
\caption{Performance of various instruct models on HumanEval, MBPP, LiveCodeBench, and the ``instruct'' task of BigCodeBench subset. Our finetuned models' performances are in the highlighted rows of the table. The best performances are marked in bold.
}
\label{tab:main}
\end{table}

\section{Analyses and Findings}

\subsection{Effectiveness of LLM-based Filtering and Verification}
\begin{table}[ht]
\centering
\resizebox{1.0\linewidth}{!} {%
\def\arraystretch{1.2}%
{\setlength{\tabcolsep}{5pt}
\begin{NiceTabular}{l|c|c|c|c c|c c}
\toprule
\multirow{2}{*}{\bf Data Selection Criteria} & \multirowcell{2}{\bf Data \\ \bf Size} & \multirowcell{2}{\bf Execution \\ \bf Pass Rate} & \multirowcell{2}{\bf Assessment \\ \bf Score} & \multicolumn{2}{c|}{\bf HumanEval} & \multicolumn{2}{c}{\bf MBPP} \\
& & & & HE & HE+ & MBPP & MBPP+ \\ 
\midrule
Random selection                    & 500k & 72.4\% & 4.42 & 82.9 & 77.8 & 81.0 & 70.1 \\
UTE Failures                        & 500k & 0\% & 4.22 & 80.0 & 75.3 & 80.1 & 69.8 \\
UTE Passes                          & 500k & 100\% & 4.53 & 83.1 & 78.4 & 81.4 & 70.4 \\
LLM Judgment Score = 5.0            & 500k & 77.4\% & 5.0 & 84.8 & 80.5 & 82.3 & 71.4 \\
\bottomrule
\end{NiceTabular}
}
}
\caption{Evaluation results demonstrating the effectiveness of filtering based on unit-test execution (UTE) feedback and LLM judgment scores to finetune Qwen2.5-Coder-7B model.
}
\label{tab:exec_filter_result}
\end{table}

We perform an ablation study to determine the effectiveness of filtering the instructions based on the synthetic unit test generation (\autoref{sec:test_case_generation}) and also the response quality assessments (\autoref{sec:response_quality}) done by LLMs. We randomly selected 500k samples from our \ourdataset dataset and compared this to 500k samples filtered by generated test cases and LLM-as-a-judge. Selecting the question-solution pairs that pass the generated test cases clearly outperforms those that failed all the test cases and marginally improves results compared to random selection. However, LLM-as-a-judge performs better than all other cases and is the most suitable verifier in our dataset. Additionally, we can observe a correlation between execution pass rate and judgment scores.

Prior works have found notable success with using test case generation to filter solutions \cite{wei2024selfcodealign}. Our filtering differs in that we are not filtering by selecting the best solution to the same problem but instead filtering out question-solution pairs. This means there is a tradeoff between diversity and correctness where we filter out unique questions that may have correct solutions but perform poorly in test case execution. This explains why unit-test execution performs only marginally better than random while LLM-as-a-judge further improves results, as it is agnostic to the indirect executability of the code. We include the details from test case generation and LLM-as-a-judge verification for all 5M samples in \ourdataset and encourage future work to further explore verification methods.


\begin{figure}[ht]
  \centering
  \includegraphics[width=0.55\textwidth]{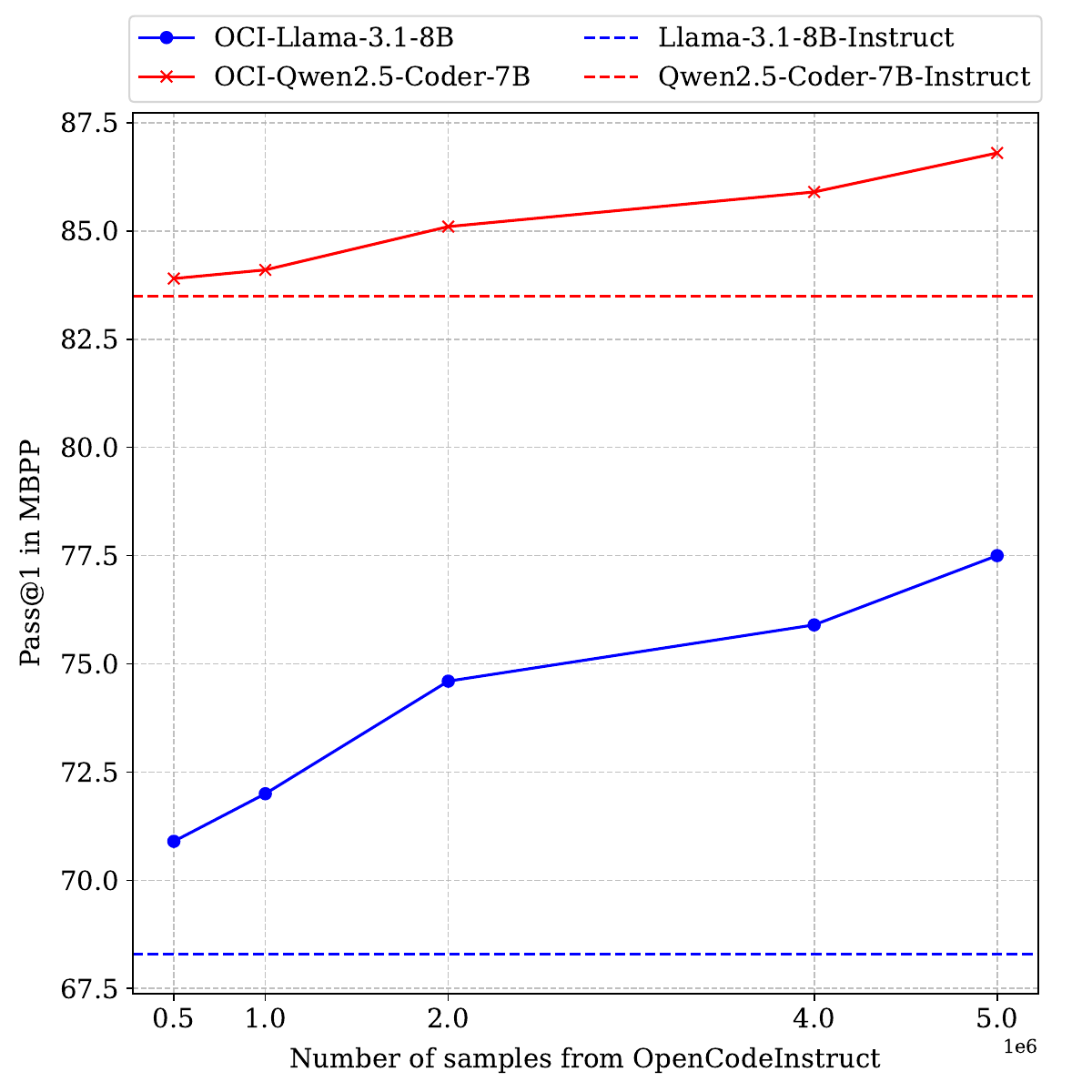}
  \caption{Finetuned model performances (Pass@1) on MBPP tasks when finetuned with different number of samples from \ourdataset.}
  \label{plot:scaling_law}
\end{figure}

\subsection{Impact of Synthetic Data Size}
In \autoref{plot:scaling_law}, we demonstrate Pass@1 score on MBPP benchmark with respect to increasing amounts of the \ourdataset samples. Notably, fine-tuning Qwen2.5-Coder-7B-Base and Llama-3.1-8B-Base with just 500k samples already surpasses their respective instruct-tuned versions. Performance consistently improves with increasing sample size, peaking at 5 million samples, which is the full size of \ourdataset. As noted earlier, Llama3, being a general-purpose LLM, experiences more significant performance gains across all sample sizes.

\subsection{Impact of Instruction Generation Algorithm}
\label{sec:impact_of_inst_algo}

Instruction generation in \textsc{Genetic-Instruct} is mainly based on two generation algorithms (\textsc{self-instruct} and \textsc{evol-instruct}). We performed an ablation on the effect of each of these algorithms on synthetic instruction generation that could impact downstream code generation performance. We separated the instructions generated by \textsc{Genetic-Instruct} based on the last operation applied on them and trained individual models. As shown in \autoref{tab:main_ablation},while both algorithms yield competitive results, one performs better than the other on certain benchmarks. It shows the difference between the capability and coverage of each generation algorithm. While \textsc{self-instruct} can broaden the domain scope of the problems, \textsc{evol-instruct} is good at diversifying the problems locally by making them harder or easier. These results indicate that both algorithms contribute unique and necessary capabilities to maximize benchmark performance.

\begin{table}[t]
\centering
\resizebox{0.95\linewidth}{!} {%
\def\arraystretch{1.0}%
{\setlength{\tabcolsep}{4pt}
\begin{NiceTabular}{l|l|c|c c|c c}
\toprule
\multirow{2}{*}{\bf Model} & \multirow{2}{*}{\bf Component} & \multirowcell{2}{\bf Numbers\\ \bf of samples} & \multicolumn{2}{c|}{\bf HumanEval} & \multicolumn{2}{c}{\bf MBPP} \\
& & & HE & HE+ & MBPP & MBPP+ \\ 
\midrule
\multicolumn{7}{l}{\bf Ablation (\ref{sec:impact_of_inst_algo}): Impact of Instruction Generation Algorithm} \\ \midrule
\multirow{2}{*}{\bf OCI-Llama-3.1-8B}      & Self-Instruct & 3M & 68.3	& 65.2 & 72.0 & 63.2 \\
                                                & Evol-Instruct & 2M & 68.3 & 65.9 & 72.5 & 63.8 \\
\arrayrulecolor{gray!70} \midrule \arrayrulecolor{black}
\multirow{2}{*}{\bf OCI-Qwen2.5-Coder-7B}  & Self-Instruct & 3M & 84.1 & 78.0 & 82.3 & 71.4 \\
                                                & Evol-Instruct & 2M & 87.2 & 80.5 & 82.8 & 72.8 \\
\midrule
\multicolumn{7}{l}{\bf Ablation (\ref{sec:impact_of_seed}): Impact of Seed Population} \\ \midrule
\multirow{3}{*}{\bf OCI-Llama-3.1-8B}      & Algorithmic (S)   & 2.5M & 69.5 & 65.9 & 73.5 & 63.5 \\
                                                & Algorithmic (L)   & 2.5M & 71.3 & 69.5 & 74.1 & 64.9 \\
                                                & Generic     (L)   & 2.5M & 71.7 & 70.3 & 74.0 & 62.2 \\
\arrayrulecolor{gray!70} \midrule \arrayrulecolor{black}
\multirow{3}{*}{\bf OCI-Qwen2.5-Coder-7B}  & Algorithmic (S)   & 2.5M & 81.7 & 76.8 & 83.3 & 71.4 \\
                                                & Algorithmic (L)   & 2.5M & 84.8 & 79.3 & 83.1 & 72.2 \\
                                                & Generic     (L)   & 2.5M & 85.4 & 79.3 & 83.3 & 71.7 \\
\midrule
\multicolumn{7}{l}{\bf Ablation (\ref{sec:instruct_format}): Impact of Instruction Formatting} \\ \midrule
\multirow{2}{*}{\bf OCI-Llama-3.1-8B}      & NL-to-Code        & 2M & 70.7	& 67.1 & 74.6 & 65.6 \\
                                                & Code-to-Code      & 2M & 68.3 & 62.2 & 71.2 & 61.1 \\
\arrayrulecolor{gray!70} \midrule \arrayrulecolor{black}
\multirow{2}{*}{\bf OCI-Qwen2.5-Coder-7B}  & NL-to-Code        & 2M & 85.4	& 79.9 & 83.3 & 73.0 \\
                                                & Code-to-Code      & 2M & 80.5 & 74.4 & 81.8 & 70.1 \\
\bottomrule
\end{NiceTabular}
}
}
\caption{Evaluation results of ablation study on instruction generation algorithm, seed population types (algorithmic and generic), and scale (S: small-scale and L: large-scale).
}
\label{tab:main_ablation}
\end{table}

\subsection{Impact of Seed Population}
\label{sec:impact_of_seed}

To assess the influence of seed population on instruction quality, we performed two ablation studies. First, we generated 2.5 million synthetic samples using a smaller set of algorithmic questions from Tiger-Leetcode \citep{tigerbotkaggle2023}. We compare this synthetic dataset with a subsample from \ourdataset where TACO is used as seeds (a larger algorithmic seed set). The evaluation results depicted in \autoref{tab:main_ablation} show that a large-scale seed based instruction set results in substantial performance gains. Subsequently, we compared models trained separately on algorithmic and generic instruction subsamples from \ourdataset. Although HumanEval and MBPP showed comparable performance, the combined instruction set yielded superior results, as shown in \autoref{tab:main}. This underscores the importance of seed population characteristics, including size, domain coverage, and diversity in synthetic instruction data generation.

\subsection{Impact of Instruction Formatting: NL-to-Code vs. Code-to-Code}
\label{sec:instruct_format}

We investigated the impact of instruction format on code generation performance, contrasting Natural Language-to-Code (NL-to-Code) and Code-to-Code formats, exemplified by the two popular function-level code generation benchmarks, MBPP and HumanEval, respectively. Using Qwen2.5-32B-Instruct, we reformatted \ourdataset instructions from NL-to-Code to Code-to-Code style using few-shot prompting (see the template in \autoref{fig:he_style_prompt}). Finetuning on these formats (\autoref{tab:main_ablation}) revealed that NL-to-Code instructions significantly outperformed Code-to-Code across all benchmarks, including HumanEval. We hypothesize that NL-to-Code is a more effective learning format for LLMs.


\subsection{Code Generation with Different Models}

\begin{figure*}[ht!]
\vspace{-2mm}
\centering
\includegraphics[width=0.8\textwidth]{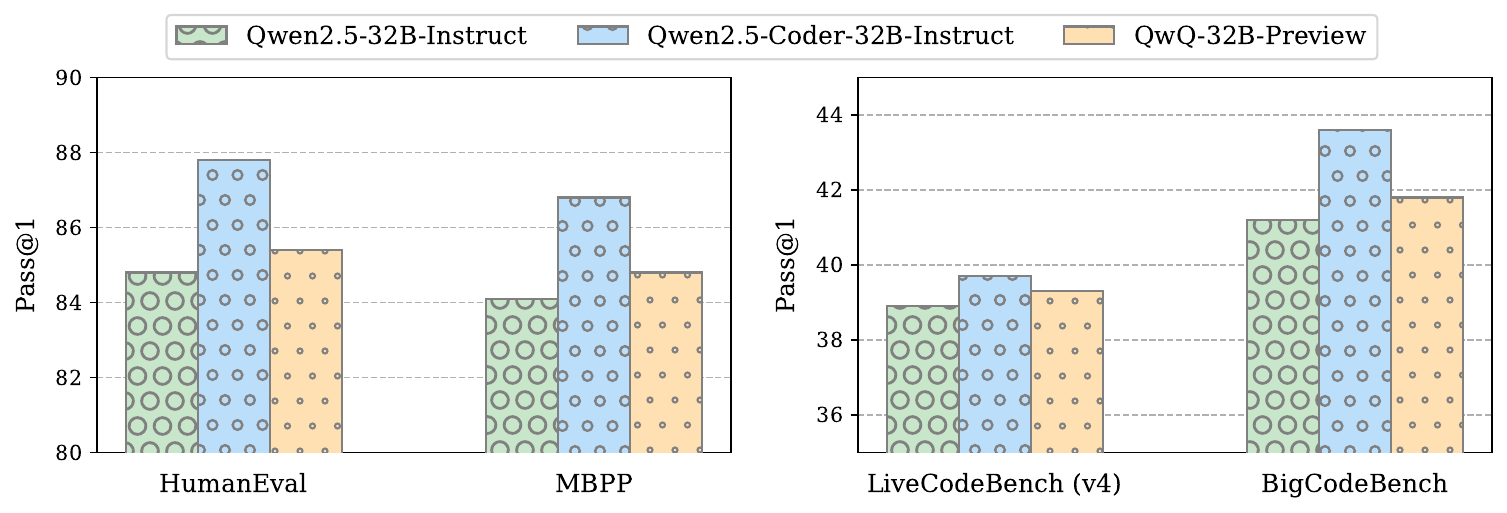}
\vspace{-2mm}
\caption{Performance comparison of finetuning \texttt{Qwen2.5-Coder-7B} using \ourdataset, across benchmarks when code solutions are generated by three different LLMs.}
\label{plot:impact_code_generator}
\vspace{-2mm}
\end{figure*}

We study the impact of using different code generation models on resultant benchmarks scores as visualized in \autoref{plot:impact_code_generator}. As expected, selecting a model that performs better at the target benchmarks also translates to improved performance when evaluating a model distilled from its generated solutions. In our case, Qwen2.5-Coder-32B-Instruct has the highest HumanEval, MBPP and BigCodeBench scores and this leads to a several point improvement over alternatives. While QwQ-32B-Preview exhibits strong reasoning and excellent benchmark results, its output length exceeds our 1024 token-generation limit. We consequently used a prefix (\verb|```python|) to enforce code-only generation, potentially sacrificing solution quality. We suggest exploring code generation incorporating reasoning traces as a direction for future research.

\subsection{\textsc{OSS-Instruct} Samples vs. \ourdataset}

\begin{table}[ht]
\centering
\def\arraystretch{1.0}%
\begin{tabular}{l |c c|c c}
\toprule
\multirow{2}{*}{\bf Model} & \multicolumn{2}{c|}{\bf HumanEval} & \multicolumn{2}{c}{\bf MBPP} \\
& HE & HE+ & MBPP & MBPP+ \\ 
\midrule
\multicolumn{5}{l}{\bf SFT w/ OSS-Instruct Samples (4M samples)} \\
\midrule
OSS-I-Llama-3.1-8B       & 69.5 & 63.4 & 70.4 & 60.8 \\
OSS-I-Qwen2.5-Coder-7B   & 82.9 & 73.8 & 84.4 & 73.3 \\
\midrule
\multicolumn{5}{l}{\bf SFT w/ OpenCodeInstruct (4M subsample)} \\
\midrule
OCI-Llama-3.1-8B       & 78.7 & 73.2 & 77.5 & 66.4 \\
OCI-Qwen2.5-Coder-7B   & 86.8 & 83.2 & 87.2 & 75.3 \\
\bottomrule
\end{tabular}
\caption{OSS-Instruct vs. OpenCodeInstruct.} 
\label{tab:oss_instruct}
\end{table}

In \autoref{tab:oss_instruct} we outline the comparison of using an equivalent 4 million samples from \textsc{OSS-Instruct} and \ourdataset. To generate the \textsc{OSS-Instruct} dataset, we repeated the data generation pipeline three times, utilizing the same 1.43 million Python functions in each run. The results presented in \autoref{tab:oss_instruct} show that finetuning Llama-3.1-8B results in 9.8 and 5.6 points improvement in HE+ and MBPP+, respectively, by upgrading to using \ourdataset. Similarly, finetuning the more capable Qwen2.5-Coder-7B leads to an improvement of 9.4 and 2.0 for HE+ and MBPP+ respectively. These findings confirm that \ourdataset offers a higher quality instruction dataset on a per sample basis alongside the added benefit of containing more samples overall.

\section{Related Work}


\paragraph{Large language models for code} Large Language Models (LLMs), trained on billions of lines of code, have shown remarkable proficiency in various software engineering tasks. This includes repository-level code generation \citep{zhang-etal-2023-repocoder, ding2023crosscodeeval, wu2024repoformer}, automated program repair \citep{xia2022less, wei2023copiloting, jiang2023impact, bouzenia2024repairagent, haque2023fixeval}, performance optimization \citep{cummins2023large}, code translation \citep{lachaux2020unsupervised, pan2023understanding, ahmad-etal-2023-avatar, ahmad-etal-2023-summarize}, and software testing  \citep{xia2023keep, deng2023large, yuan2023no, schafer2023empirical, lemieux2023codamosa}. Core models like PLBart \citep{ahmad-etal-2021-unified}, CodeT5 \citep{wang-etal-2021-codet5}, CodeGen \cite{nijkamp2023codegen}, StarCoder \citep{li2023starcoder, lozhkov2024starcoder}, Code Llama \citep{roziere2023code}, and DeepSeek-Coder \citep{guo2024deepseek} are pre-trained on massive codebases, providing a strong foundation for general code generation and comprehension. Recent advancements focuses on fine-tuning \citep{luo2024wizardcoder} and prompt engineering \citep{chen2023teaching} to specialize these models for specific coding challenges.

\paragraph{Instruction tuning with synthetic data}
Instruction tuning aims to improve large language models (LLMs) by fine-tuning them on instruction-response pairs \citep{wei2022finetuned}. Recognizing the difficulty of acquiring high-quality instructional data, researchers have increasingly focused on synthetic data generation. \textsc{Self-Instruct} \citep{wang-etal-2023-self-instruct} pioneered this approach, utilizing a foundation LLM to generate instruction-response pairs for its own fine-tuning. Building upon this, WizardLM \citep{xu2024wizardlm} and WizardCoder \citep{luo2024wizardcoder} introduced \textsc{Evol-Instruct} and \textsc{Code Evol-Instruct}, respectively, employing heuristic prompts to enhance data complexity and diversity. \citet{majumdar2024genetic} draws inspiration from evolutionary processes to create a scalable method for synthetic data generation.
In concurrent works, \textsc{OSS-Instruct} \citep{wei2023magicoder} and \textsc{Reverse-Instruct} \citep{wu2024inversecoder} shifted towards leveraging real code snippets as a data source. \textsc{SelfCodeAlign} \citep{wei2024selfcodealign} further refines synthetic data generation through self-alignment, where a base code LLM generates data for its own instruction fine-tuning.

\section{Conclusion}
We present \ourdataset, the largest LLM-generated code instruction tuning dataset to date. Fine-tuning Llama3 and Qwen2.5-Coder across various model sizes with \ourdataset significantly outperforms their instruction-tuned counterparts on HumanEval, MBPP, LiveCodeBench, and BigCodeBench. We also provide insights into the effectiveness of design choices within the \ourdataset pipeline, demonstrating their impact on downstream code generation tasks. The \ourdataset dataset will be fully open-sourced to facilitate future LLM-for-code research.



\bibliography{bib/anthology,bib/custom}
\bibliographystyle{template/colm2025_conference}

\appendix
\clearpage
{
\centering
\Large\bf Supplementary Material: Appendices \\ [20pt]
}

\begin{figure*}[ht!]
\centering
\includegraphics[width=\textwidth]{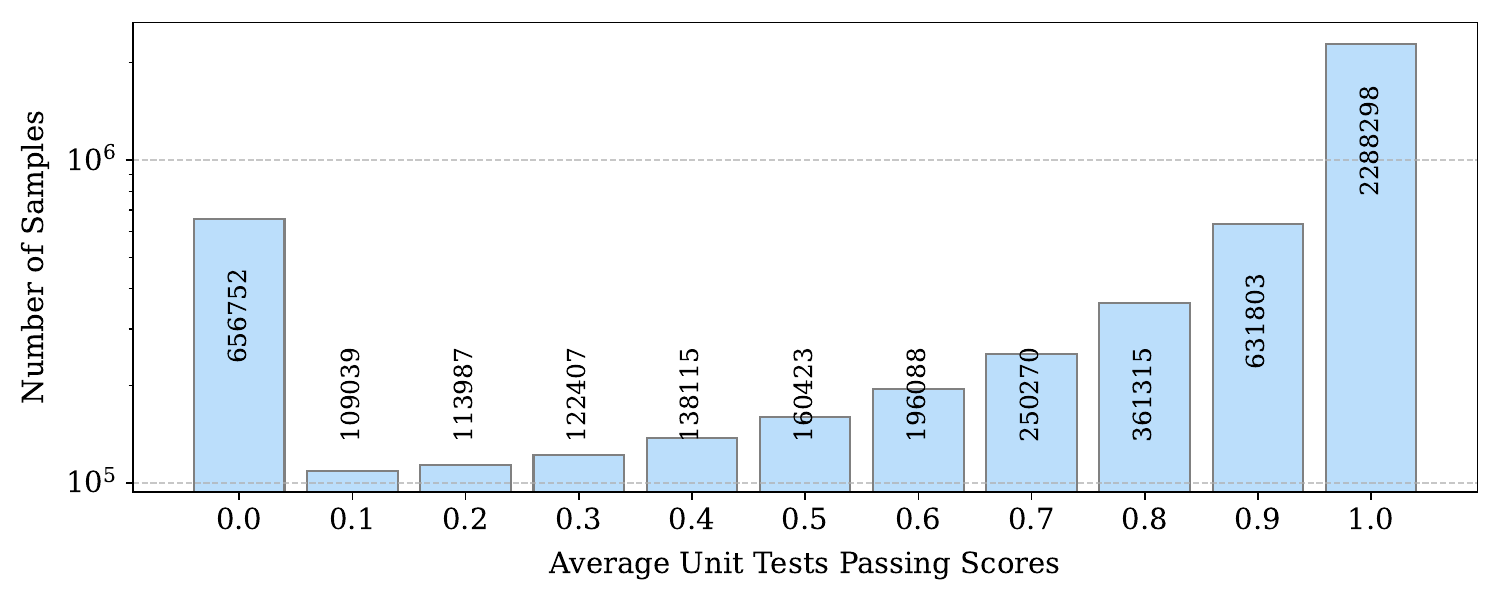}
\vspace{-5mm}
\caption{Histogram of average unit test pass rates for \ourdataset samples.}
\label{plot:ut_score_distribution}
\end{figure*}

\begin{figure*}[ht!]
\centering
\includegraphics[width=\textwidth]{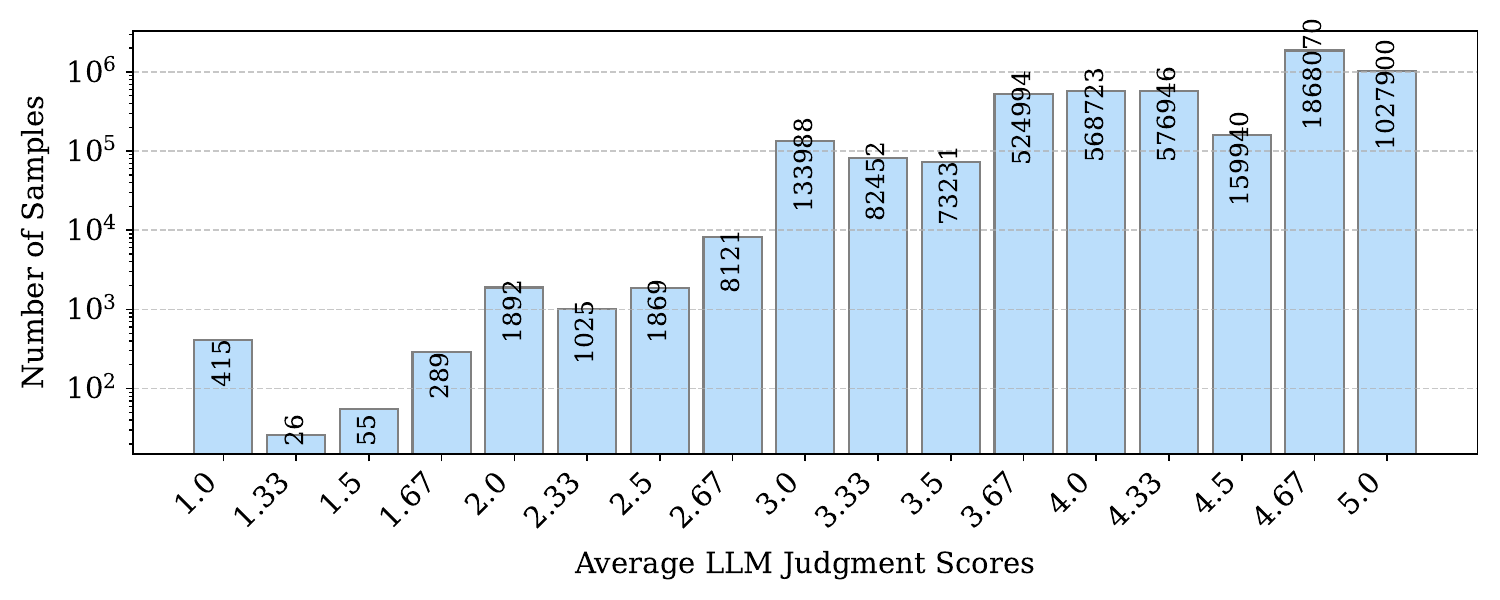}
\vspace{-5mm}
\caption{Histogram of average llm-as-a-judge scores for \ourdataset samples.}
\label{plot:llm_score_distribution}
\end{figure*}

\begin{figure*}[ht!]
    \centering
    \includegraphics[width=0.75\textwidth]{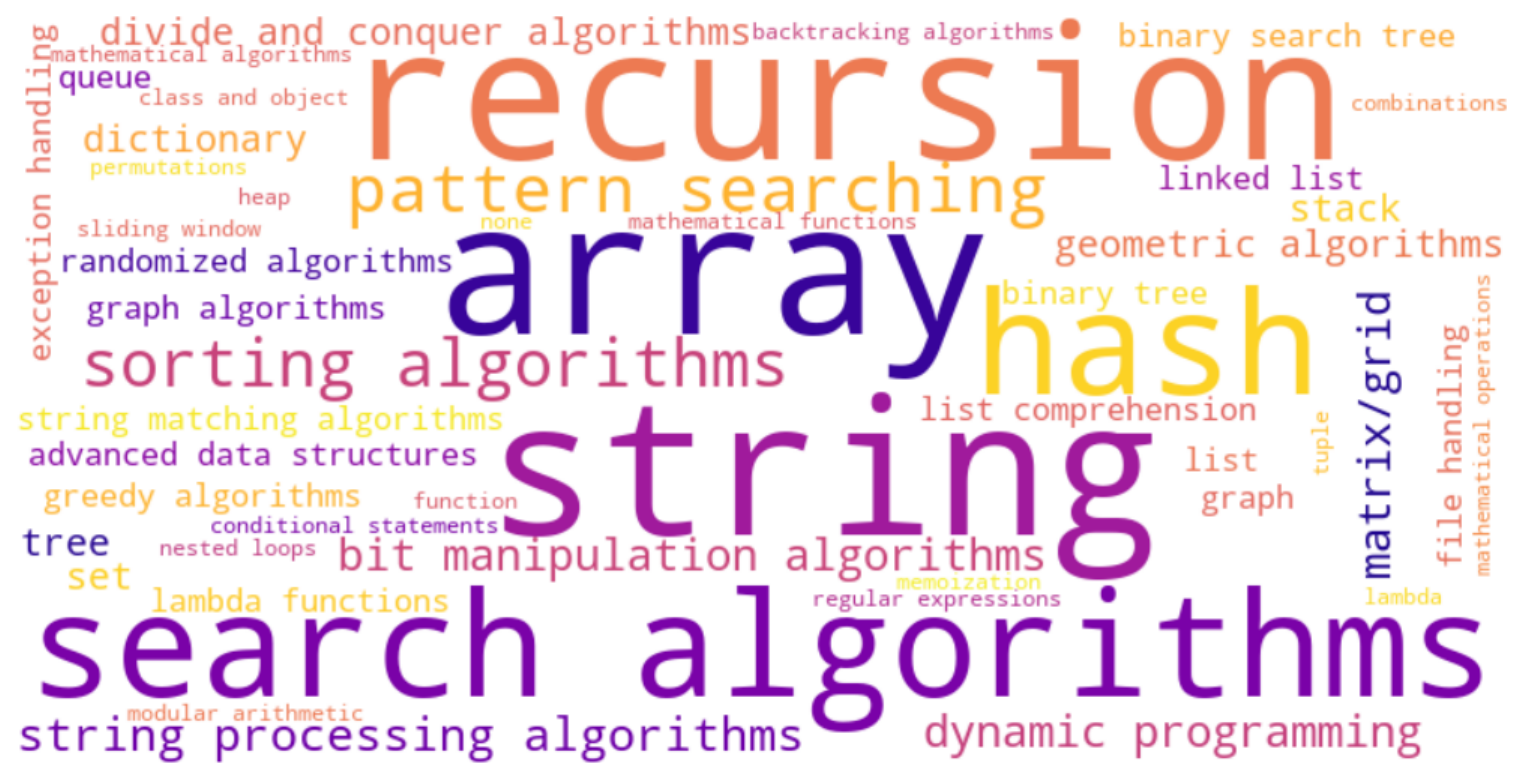}
    \caption{Visualization of LLM generated skills that are used or demonstrated in code solutions in \ourdataset.}
    \label{plot:skills}
\end{figure*}

\begin{figure*}[ht!]
\centering

\begin{tcolorbox}[title={Test Case Generation Prompt}, colback=red!0, left=2pt,right=2pt,top=0pt,bottom=0pt]

{ 
You are an expert at writing assertion test cases and below is a question with function signature and completed code solution. You must generate 10 assert statements that will be used to evaluate the code solution's correctness which may or may not be correct. Here are some examples that you should use as a reference:

\vspace{0.15cm}
Question:
\begin{lstlisting}
from typing import Optional
def first\_repeated\_char(s: str) -> Optional[str]:
    """ 
    Find the first repeated character in a given string.
    >>> first\_repeated\_char("abbac")
    'a'
    """
\end{lstlisting}

\vspace{0.15cm}
Solution:
\begin{lstlisting}
from typing import Optional
def first_repeated_char(s: str) -> Optional[str]:
    """ 
    Find the first repeated character in a given string.
    >>> first_repeated_char("abbac")
    'a'
    """
    for index, c in enumerate(s):
        if s[:index + 1].count(c) > 1:
            return c
    return None
\end{lstlisting}

\vspace{0.15cm}
Test Cases:
\begin{lstlisting}
<assertion>assert first_repeated_char("!@#$%^&*!") == "!"</assertion>
<assertion>assert first_repeated_char("abcdedcba") == "d"</assertion>
<assertion>assert first_repeated_char("") == "None"</assertion>
<assertion>assert first_repeated_char("aaaa") == "a"</assertion>
<assertion>assert first_repeated_char("a") == "None"</assertion>
\end{lstlisting}

Here are guidelines for writing the assertion test cases:
\begin{enumerate}[leftmargin=*, itemsep=0pt]
    \item You must wrap each assertion test case with tags \opentag{assertion} and \closetag{assertion}.
    \item Do not start the assert with any indents or spaces.
    \item You must not import any unit testing libraries for the assertions such as ``unittest'' or ``pytest''.
    \item Each assertion must be complete and immediately executable. Assume the code solution is provided, do not repeat it.
    \item Avoid unnecessary string literals, incorrect escaping, wrapping in \lstinline{```python} or other redundancies.
    \item Remember, it is your responsibility to carefully read the question and generate test cases that will evaluate the correctness of the solution.
\end{enumerate}

\vspace{0.3cm}
Here is the question and code solution you must provide assertion test cases for:

\vspace{0.3cm}
Question: \\
\{question\}
\vspace{0.3cm}

Solution: \\
\{solution\}
\vspace{0.3cm}

Test Cases:
}
\end{tcolorbox}

\caption{Prompt template for test case generation.}
\label{fig:utg_prompt}
\end{figure*}

\begin{figure*}[ht!]
\centering

\begin{tcolorbox}[title={HumanEval Tasks Style Instruction Generation Prompt}, colback=red!0, left=2pt,right=2pt,top=0pt,bottom=0pt]

{ 
Take the following examples of function signatures as a reference.

\vspace{0.2cm}
Example1:
\begin{lstlisting}
def string_to_md5(text):
    """
    Given a string 'text', return its md5 hash equivalent string.
    If 'text' is an empty string, return None.
    
    >>> string_to_md5('Hello world') == '3e25960a79dbc69b674cd4ec67a72c62'
    """
\end{lstlisting}

\vspace{0.2cm}
Example2:
\begin{lstlisting}
def generate_integers(a, b):
    """
    Given two positive integers a and b, return the even digits between a
    and b, in ascending order.
    
    For example:
    generate_integers(2, 8) => [2, 4, 6, 8]
    generate_integers(8, 2) => [2, 4, 6, 8]
    generate_integers(10, 14) => []
    """
\end{lstlisting}

\vspace{0.2cm}
Now, generate a function signature for the following question and solution. Use the above mentioned examples as a reference.

\vspace{0.2cm}
Question: \\
\{question\}

\vspace{0.2cm}
Solution: \\
\{solution\}

\vspace{0.2cm}
Note that, in the generated function signature, function body should be empty (do not even write pass statement).

}
\end{tcolorbox}
\caption{Prompt template for HumanEval tasks style instruction generation.}
\label{fig:he_style_prompt}
\end{figure*}

\begin{figure*}[htbp]
\centering

\begin{tcolorbox}[title={Judge LLM Prompt}, colback=red!0, left=2pt,right=2pt,top=0pt,bottom=0pt]

{ 
You are an expert in evaluating coding questions and solutions. You are given the following rubric to evaluate the code solution which may or may not be correct.

\vspace{0.2cm}
Programming Solution Evaluation Rubric (Scale 1-5)

\vspace{0.2cm}
\#\# Requirement Conformance:
\begin{enumerate}[leftmargin=*, itemsep=0pt]
    \item Ignores most specifications.
    \item Addresses few requirements.
    \item Meets basic requirements but misses some details.
    \item Addresses most requirements with minor gaps.
    \item Fully meets or exceeds all specified requirements.
\end{enumerate}

\vspace{0.2cm}
\#\# Logical Correctness:
\begin{enumerate}[leftmargin=*, itemsep=0pt]
    \item Fundamental logic is flawed.
    \item Major logical errors present.
    \item Mostly correct with some minor issues.
    \item Largely correct and consistent logic.
    \item Completely correct and optimally structured.
\end{enumerate}

\vspace{0.2cm}
\#\# Edge Case Consideration:
\begin{enumerate}[leftmargin=*, itemsep=0pt]
    \item No edge cases considered.
    \item Minimal consideration of unusual inputs.
    \item Some edge cases addressed but not all.
    \item Most edge cases are anticipated and handled.
    \item Comprehensive and robust handling of all potential edge cases.
\end{enumerate}

\vspace{0.2cm}
You have to provide scores for each criterion and justification for your score as a JSON response as follows. \\
\verb|```|json
\begin{lstlisting}
{
  "requirement_conformance": {
    "score": [1-5],
    "justification": "reasoning for scoring on requirement conformance"
  },
  "logical_correctness": {
    "score": [1-5],
    "justification": "reasoning for scoring on logical correctness"
  },
  "edge_case_consideration": {
    "score": [1-5],
    "justification": "reasoning for scoring on edge case consideration"
  }
}
\end{lstlisting}
\verb|```|
  
\vspace{0.2cm}
Now evaluate the question and code solution using the above mentioned rubric. Don't generate anything except the JSON response.

\vspace{0.2cm}
Question: \\
\{question\}
\vspace{0.2cm}

Solution: \\
\{solution\}
}
\end{tcolorbox}
\caption{Prompt template for an LLM to function as a Judge in evaluating code solutions for corresponding coding tasks.}
\label{fig:llm_judge_prompt}
\end{figure*}

\begin{figure*}[ht!]
\centering

\begin{tcolorbox}[title={Code to Skills Generation Prompt}, colback=red!0, left=2pt,right=2pt,top=0pt,bottom=0pt]

{ 
You are an expert in providing data structure and algorithm skills used/demonstrated in Python code. You are given the following list.

\vspace{0.15cm}
A list of data structure skills:
\begin{enumerate}[leftmargin=*, itemsep=0pt]
    \item Array
    \item Matrix/Grid
    \item String
    \item Stack
    \item Queue
    \item Linked list
    \item Hash
    \item Tree
    \item Binary Tree
    \item Binary Search Tree
    \item Heap
    \item Graph
    \item Advanced Data Structures
\end{enumerate}

\vspace{0.15cm}
A list of algorithm skills:
\begin{enumerate}[leftmargin=*, itemsep=0pt]
    \item Search algorithms
    \item Sorting algorithms
    \item Graph algorithms
    \item Greedy algorithms
    \item Backtracking algorithms
    \item Divide and conquer algorithms
    \item Recursion
    \item Dynamic programming
    \item Pattern searching
    \item Geometric algorithms
    \item Branch and bound algorithms
    \item Randomized algorithms
    \item Bit manipulation algorithms
    \item String matching algorithms
    \item String processing algorithms
\end{enumerate}

\vspace{0.15cm}
Now, given the following Python code snippet, generate a list of top 3 skills that are demonstrated or required to understand and work with the code.

Solution: \\
\{solution\}

\vspace{0.15cm}
Guidelines for generating the skills:
\begin{enumerate}[leftmargin=*, itemsep=0pt]
    \item Please provide the skills as a list of strings in Python format.
    \item If none of the listed skills are relevant, generate an empty list.
    \item Don't provide any explanation.
\end{enumerate}

}
\end{tcolorbox}
\caption{Prompt template for Code to Skills generation.}
\label{fig:skills_prompt}
\end{figure*}

\begin{figure}[ht!]
\centering
\includegraphics[scale=0.5]{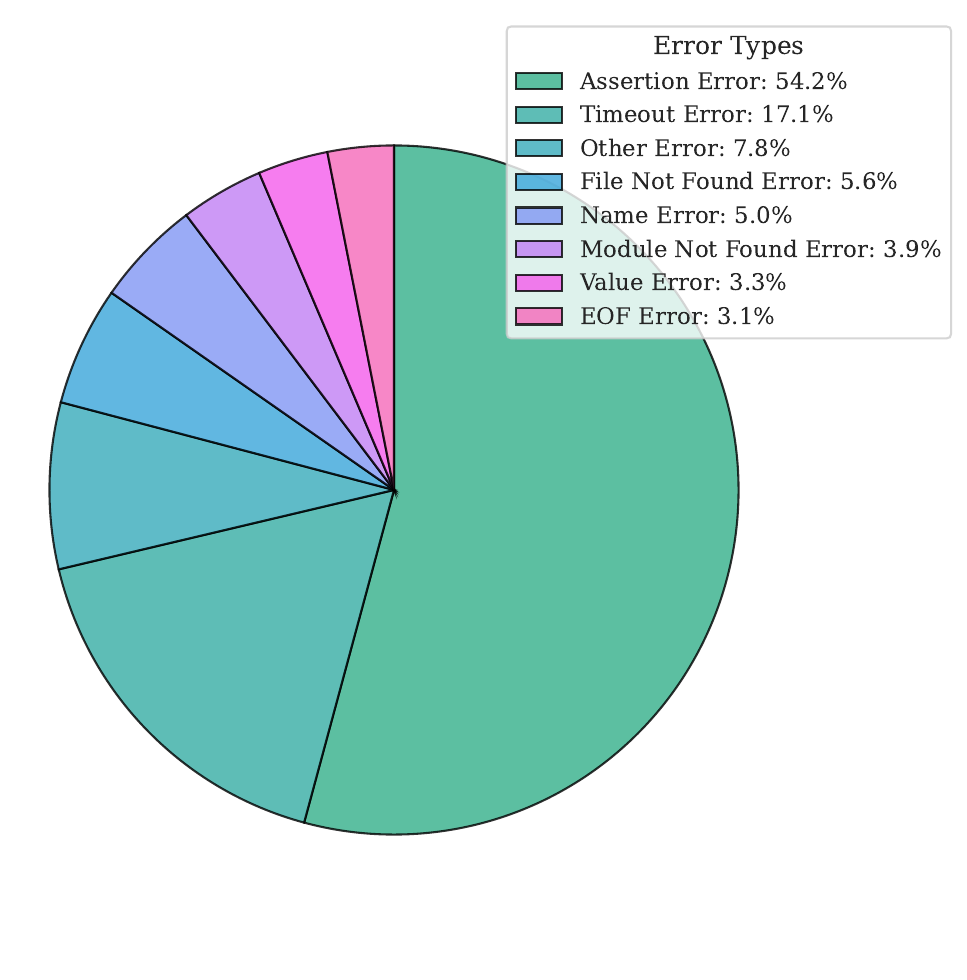}
\vspace{-10mm}
\caption{Fraction of error types in failed generated test cases.}
\label{plot:oci_error_type_distribution}
\end{figure}

\begin{figure*}[ht!]
\centering
\includegraphics[width=\textwidth]{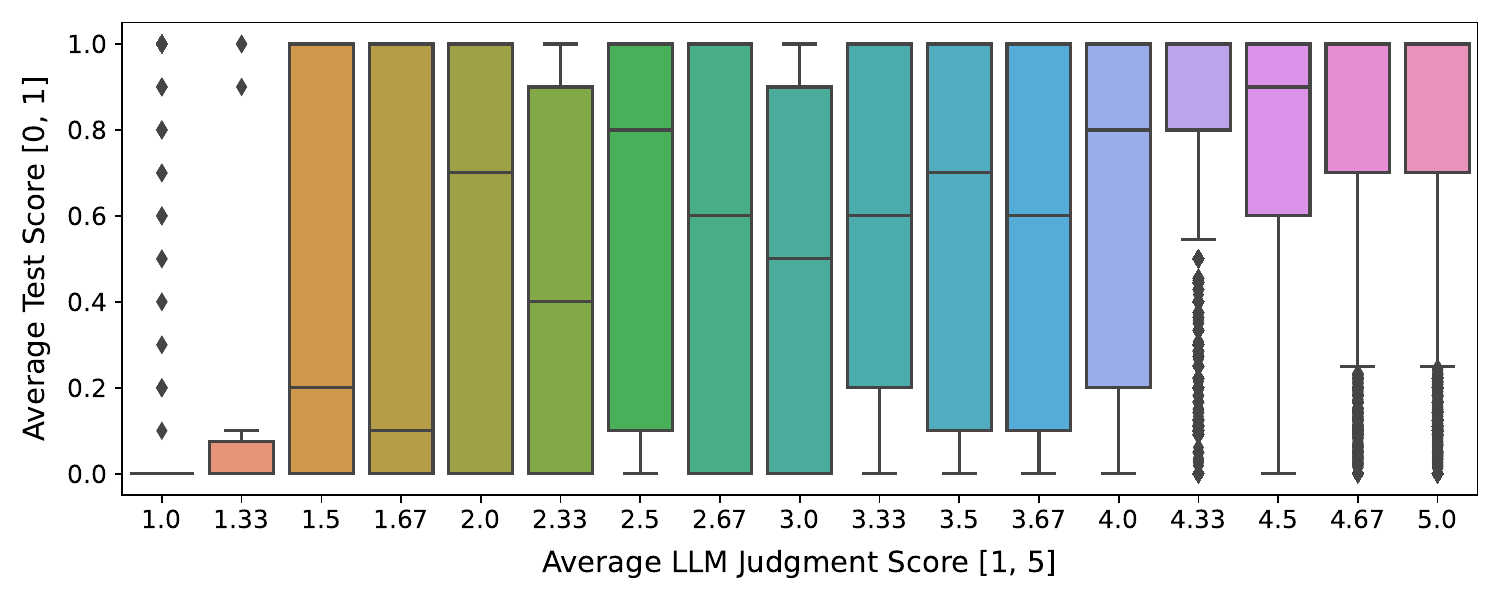}
\vspace{-5mm}
\caption{Visualization of the relationship between average unit test scores and average LLM judgment scores. The plot displays the distribution of test scores within each LLM score category, highlighting potential outliers and trends in data quality assessment.}
\label{plot:llm_vs_ute_scores}
\end{figure*}

\end{document}